\documentclass{jltp}

\usepackage{H-FromRevTeX} 
\usepackage{epsfig}    

\title{Superconductivity in Multiple Interface Geometry: Applicability
of Quasiclassical Theory}

\author{M. Ozana and A. Shelankov$^*$}
\address{
Department of Theoretical Physics, Ume{\aa} University, 901 87
Ume{\aa}, Sweden\\
${}^*$Also at A.F. Ioffe Physico-Technical Institute, 194021 St. Petersburg,
Russia.
} 
\runninghead{M. Ozana and A. Shelankov}
{Applicability of the quasiclassical theory ...}

\newcommand{\gR}{\hat{g}^{R}}
\def\baroverletter#1{\setbox1=\hbox{$#1$}
  \dimen1=\ht1
    \advance\dimen1 by 1pt
  \dimen2=\wd1
    \advance\dimen2 by -1pt
  \rlap{\hspace{.5pt}\rule[\dimen1]
              {\dimen2}{.35pt}}\box1} 
\def\ops{{\baroverletter\psi}}

\def\ofi{{\baroverletter\phi}}

\newcommand{\Hrow}[2]{\left(
\begin{array}{lcr}
#1   &,& #2
\end{array}
\right)
}
\newcommand{\Hcolumn}[2]{\left(
\begin{array}{c}
#1 \\
#2
\end{array}
\right)
}
\begin{document}

\maketitle

\begin{abstract}
The method of two-point quasiclassical Green's function is reviewed
and its applicability  for description of  multiple
reflections/transmissions in layered structures is discussed.
The Green's function of a sandwich built of superconducting layers
with  a semi-transparent interface is found with the help of recently
suggested quasiclassical method [A. Shelankov, and M. Ozana,
 Phys.\ Rev.\ {\bf B 61}, 7077 (2000)], as well as exactly, from the
Gor'kov equation. 
By the comparison of the results of the two
approaches, the validity of the quasiclassical method for the description
of real  (non-integrable) systems is confirmed.
\end{abstract}

\section{INTRODUCTION}

One of the most efficient tools in the superconductivity theory is the
method of the quasiclassical Green's
function\cite{Eil68,LarOvc,Sch81,SerRai83}.  In recent decades, the
quasiclassical technique has been successfully applied to a broad
variety problems.  Efficiency of the quasiclassical theory is due to
the fact that it allows one to get description directly on the large
spatial and temporal scales relevant for superconductivity, completely
eliminating from the theory short ``quantum'' scales given by the
Fermi wave length.  Ignoring fine features on the quantum scale, the
theory gives a closed set of equations for smoothly varying envelops.

The quasiclassical approximation is applicable in
superconductors because 
the order parameter $\Delta $ varies on 
the
coherence length $\xi \sim v_{F}/ \Delta $,
 which usually much exceeds the Fermi wave length
$\lambdabar_{F} \sim \hbar / p_{F}$,  $v_{F}$ and $p_{F}$ being the Fermi
velocity and momentum, respectively. As first discussed by Andreev\cite{And64}, the slow spatial modulations of the order parameter   
cannot significantly change the momentum. Then, the two-component wave
function $\psi $  can be presented as 
$\psi = e^{{i\over \hbar } {\bf p}_{F}
\bf{\cdot r}} {u \choose v}$ where $u$ and $v$ are  functions which slowly
vary along  a straight line, a classical trajectory corresponding to the  
momentum ${\bf p}_{F}$\cite{And64}. 
After the factorization, the
Fermi wave length  does not enter the theory anymore.

The quasiclassical condition is violated if the potential energy
varies on a scale of order of $\lambdabar_{F}$ e.g. when
disorder (impurities) or interfaces are present.  Disorder does not pose any
difficulty as long as one is interested only in the impurity averaged
properties: On average, disorder enters the equations via the
self-energy, acting like a smooth effective potential.  Scattering by
an isolated interface also does not create conceptional difficulties to
the quasiclassical theory: one can include the reflection and
transmission events by an appropriate boundary condition for the
Andreev wave function~\cite{She80b} or the Green's function
\cite{Zai84}. 
It is worthwhile to notice at this point that the quasiclassical scheme
does {\em not} give correct results for quantities local in the
$\bbox{p}- \bbox{r}$ phase space. 
For instance, a particle with the momentum $k_{Fx}$ reflected from a wall creates the density,
$|\Psi |^{2}= |e^{ik_{Fx}x} + r e^{-ik_{Fx}x}|^{2}$, which oscillates on
the Fermi wave length due to the interference term $2 {\rm Re}\;r
e^{-2ik_{F}x}$. However, the interference pattern, which is
 very sensitive to
the value of $k_{Fx}$, {\it i.e.}  the direction of incidence,
disappears after integration in a small region of the incidence
angle. 
Although the quasiclassical technique ignores the interference of the
incident and reflected wave, it
gives correct predictions for coarse-grained averages: Observables like
e.g. the electric current density, are given by the integral over the
momentum space, and the low resolution  description suffices. 

\begin{figure}[!] 
\centerline{
\epsfig{file=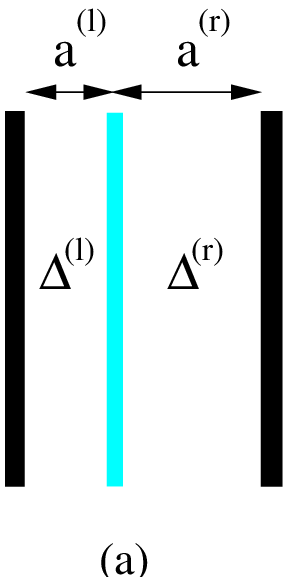,height=140pt}  \hspace{70pt}
\epsfig{file=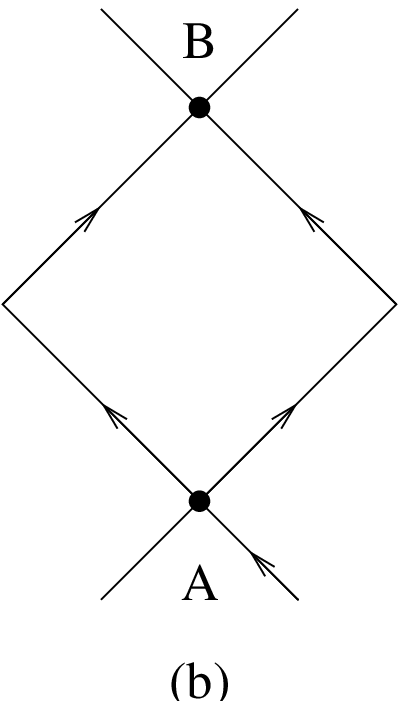,height=140pt}
}
\caption{
(a) Sandwich build of two
superconductor layers with the order parameters $\Delta^{(l,r)}$ and
thicknesses $a^{(l,r)}$ separated by  a partially transparent
interface. 
(b) The semiclassical wave which corresponds to a particle hitting the
interface at the point (knot) $A$, is split by the interface into the
left and right beams.  If the left and right layers have the same
thickness, the beams meet each other again at the knot $B$ after the
reflection by the outer walls. 
}
\label{sandwich}
\end{figure}

In many interesting cases such as
domains in the high-T$_{c}$ oxides or artificial mesoscopic structures, 
a few partially reflecting interfaces are present.
In a multi-layer geometry, the quasiclassical method meets
severe difficulties as noted in Ref.\ \onlinecite{AshAoyHar89}.
The purpose of this paper is to identify the physics behind
the difficulties and check the validity of the recently suggested
quasiclassical scheme by a numeric comparison with exact results for a
double-layer structure.

We attribute the difficulties in the theory to
the existence of interfering paths formed by sequential reflections in a
multi-layer structure.  
Consider, for instance,  a sandwich built of two superconductors, left
and right, of 
finite thicknesses, $a^{(l,r)}$ with a partially transparent interface,
Fig.\ \ref{sandwich}(a).  
We assume that the
outer walls and the interface are ideal so that all the reflections
and transmissions are specular.  A fragment of a classical trajectory
for a symmetric sandwich is shown in Fig.\ \ref{sandwich}(b).

\begin{figure}[h] 
\centerline{
\epsfig{file=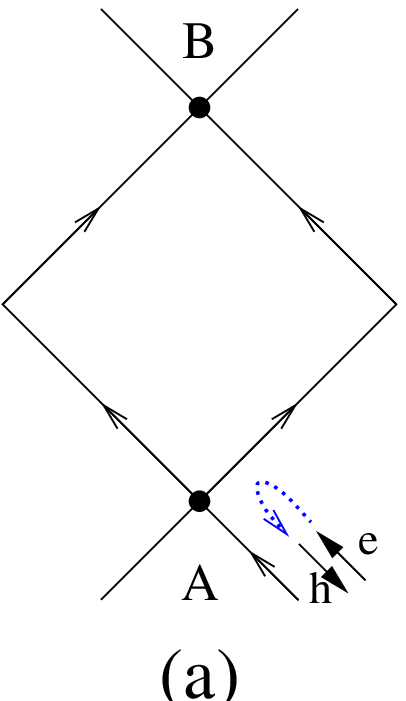,height=140pt}  \hspace{40pt}
\epsfig{file=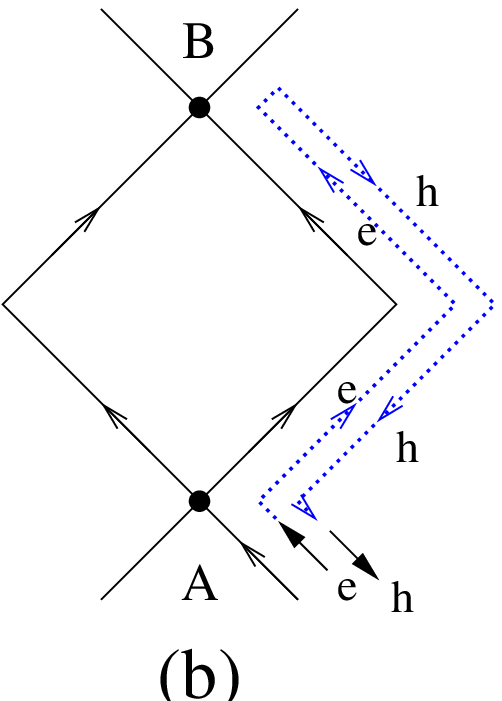,height=140pt}  \hspace{40pt}
\epsfig{file=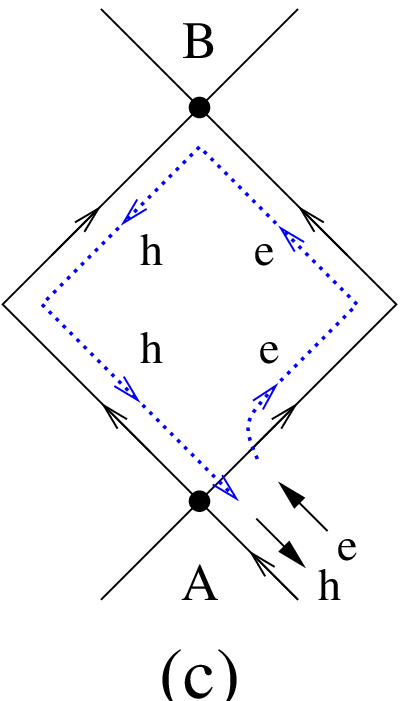,height=140pt}
}
\caption{(a) and (b)
Examples of processes included into quasiclassical theory. An electron
is Andreev reflected back as a hole and follows the original
trajectory. The loops are not formed in these cases.
(c) The simplest example of a loop-forming process.
An electron is scattered in the right arm and then
Andreev reflected as a hole in the left arm. Such a process is not
included in the quasiclassical approach which considers only tree-like
(simply connected) trajectories.
}
\label{andreevABC}
\end{figure}

Various processes of the Andreev reflection of an electron-like
excitation coming on the knot $A$ are shown in Fig.\ \ref{andreevABC}.
In the present context, the most
interesting is the process shown in Fig.\ \ref{andreevABC}(c) where the
particle makes a loop. 
The amplitude of the
process, unlike the processes in 
Fig.\ \ref{andreevABC}(a),(b)
contains the phase shift $\exp[i p_{F}({\cal L}^{(r)}-
{\cal 
L}^{(l)})]$ where ${\cal L}^{(r)}$ and ${\cal L}^{(l)}$ are the
lengths of the left and right arms, respectively.  As a result,
the total amplitude given by the sum of various contributions,
non-trivially depends on the value $p_{F}$. One comes to the
conclusion that
the Andreev factorisation, $\psi = e^{{i\over \hbar } {\bf p}_{F}
\bf{\cdot r}} {u \choose v}$, which is the essence of the
quasiclassical method, becomes impossible.

It is clear that the loop contribution is very sensitive to
geometry and surface roughness. Indeed, (independent) variations of ${\cal
L}^{(l,r)}$ on the scale of $\lambdabar_{F}$  destroy the interference.
Besides,  loops exist only if the trajectories which are split by
knot $A$ in Fig.\ref{sandwich} meet again on knot $B$. This may happen
only  in an ideal structure; any imperfection on a scale $\gtrsim
\lambdabar_{F}$ would not allow this ``accidental'' event to happen. 
(In other words, any real physical system is non-integrable.)
In a real system,
 multiple
reflections/transmissions lead to the formation of  single-connected
tree-like trajectories as has been discussed in Ref.\ \onlinecite{SheOza00}. 
The quasiclassical technique is expected to be applicable if one
solves the equations on the tree rather than on straight line
trajectories. 

Below, we compare exact and quasiclassical theories.  For
definiteness, we consider the trajectory resolved density of states
(DOS) obtained by both approaches. It is clear from the very
beginning, that agreement may be present only on a low resolution
level: strong microscopic variations of the exact DOS in the
$\bbox{p}-\bbox{r}$ space must be smoothen by integration in a small
region of the phase space.  Even then, disagreement is expected. The
point is that the exact calculations are manageable only for an ideal
geometry whereas the quasiclassical theory is expected to be valid
only when some roughness is present. In the ideal geometry, there are
always loops which survive averaging in the phase space, and their
contribution may spoil the agreement.

The paper is organised as follows. In Sec.\ref{quasi} we review the
quasiclassical approach to the multi-layer systems\cite{SheOza00}. We
introduce the trajectory Green's function and its factorization. 
The Green's function in the double layer system is found using both
quasiclassical theory (Sec.\ref{sec:quasiGreen}) as well as Gor'kov
equations (Sec.\ref{sec:exactGreen}).
In Sec.\ref{sec:sandwich} the exact approach 
is compared with the quasiclassical theory.
The applicability and the limits of quasiclassics are
discussed.  The results are summarised in Sec.\ref{conclusions}.

\section{QUASICLASSICAL TECHNIQUE} \label{quasi}

After Eilenberger\cite{Eil68}, the quasiclassical Green's (retarded)  
function $\hat{g}_{\bbox{n}}^{R}(\bbox{r})$,
which is a $2 \times 2$ matrix in a spin-singlet Cooper pair
superconductor, is introduced as
\begin{equation}
\hat{g}_{\bbox{n}}^{R}(\bbox{r}) = {i \over \pi} \int d \xi
\hat{G}_{\bbox{p}}^{R}(\bbox{r}) \; ,
\label{i3b}
\end{equation}
where $\hat{G}_{\bbox{p}}(\bbox{r})$ is the Green's function
$\hat{G}(\bbox{r}_{1}, \bbox{r}_{2})$ in the Wigner representation
\begin{equation}
\hat{G}_{\bbox{p}}^{R}(\bbox{r}) = \int
\hat{G}^{R}(\bbox{r}+ {\bbox{ \rho }\over 2}, \bbox{r} - {\bbox{ \rho
}\over 2}) e^{i \bbox{p\cdot \rho}} \; .
\label{j3b}
\end{equation}
 
The Green's function  for a given direction of the momentum
$\bbox{n}$  obeys a differential equation which couples the
spatial points along a straight line parallel to $\bbox{n}$
(or, more precisely parallel to the Fermi velocity $\bbox{v}_{F}$ corresponding
the selected point of the Fermi surface in case of anisotropic spectrum).
The line is the classical trajectory of the electron (hole).  

Alternatively, to Eq.\ (\ref{i3b}), one may use the
formulation of the quasiclassical
technique in terms of the 2-point Green's function on classical
trajectories\cite{She85,SheOza00}.
The 2-point Green's function obeys 
the following equations
\begin{eqnarray}
\!\!\!\!\!\!\!\!
\left( 
i v {\partial \over \partial x_{1}} +
\hat{H}_{\varepsilon,\bbox{n}}^{R}(\bbox{r}_{1})\right)
\gR_{\varepsilon}(x_{1},x_{2}|\bbox{n},\bbox{R})
\!\!\!
&=&
\!\!\!
i v  \delta(x_{1} - x_{2}) 
,\;  \bbox{r}_{1}= \bbox{R}+ x_{1}\bbox{n} \; ,
\label{vra1}\\   
\!\!\!\!\!\!\!\!\!\!\!\!
\gR_{\varepsilon}(x_{1},x_{2}|\bbox{n},\bbox{R})
\! 
\left( 
\!
-i v {\partial \over \partial x_{2}} + 
\hat{H}_{\varepsilon,\bbox{n}}^{R}(\bbox{r}_{2})\right)
\!\!\!
&=&
\!\!\!
i v  \delta(x_{1} - x_{2}) 
,\; \bbox{r}_{2}= \bbox{R}+ x_{2}\bbox{n} \; ,
\label{vra}
\end{eqnarray}
where the derivative in Eq.\ (\ref{vra}) operates backwards.  
The $2\times 2$ traceless matrix
$\hat{H}_{\varepsilon, \bbox{n}}^{R}$ reads,
\[
\hat{H}_{\varepsilon, \bbox{n}}^{R}=  
 \hat{h }_{\varepsilon, \bbox{n}}^{R}
- \hat{\Sigma }_{\varepsilon, \bbox{n}}^{R}
\;\; ,  
\]
\begin{equation}
\hat{h}_{\varepsilon, \bbox{n}}^{R}=
\left(
\begin{array}{lr}
\varepsilon - \bbox{v\cdot p}_{s}&  \Delta_{\bbox{n}}\\
-\Delta^{*}_{\bbox{n}}& -\varepsilon + \bbox{v\cdot p}_{s}
\end{array}
\right) 
\;\; , \;\;  \bbox{v}= v \bbox{n} \; ,
\label{u6a}
\end{equation}
where $\Delta_{\bbox{n}}$ is the order parameter (which may dependent on
the  direction $\bbox{n}$), and 
$\bbox{p}_{s}= - {e\over c}\bbox{A}$, $\bbox{A}$ being the vector potential, 
and $\hat{\Sigma}^{R}$ is built of
the impurity self-energy  and the part of the electron-phonon
self-energy not included to the self-consistent field $\Delta $.
(Below, we omit $\bbox{R}, \bbox{n}$ and $\varepsilon
 $ for brevity and use the notation $\gR(x_{1},x_{2})$.)

The propagator
$\gR$ tends to zero at $|x_{1}-x_{2}| \rightarrow \infty $,
and  $\gR$ is an analytic function of $\varepsilon $ in the upper
half plane for any $x_{1,2}$ including $|x_{1}-x_{2}|= \infty $.

Although the observables can be expressed via the quasiclassical
1-point Green's function ($x_{1}= x_{2}$), the 2-point Green's
function turns out to be a useful intermediate object.  It gives a
full physical description of the system in the approximation where
part of the orbital degree of freedom is treated classically (no
quantum broadening in the plane $\perp \bbox{n}$), with a complete
quantum treatment of the electron-hole degree of freedom.

It is important that the construction based on the notion of smooth
classical trajectories remains valid in the presence of disorder (or
phonons), in the standard approximation when the scattering is
included on the average via the self-energy (provided $p_{F}l \gg 1$,
$l$ being the mean free path).

\subsection{Constructing 2-point Green's function}

The two-point matrix Green's function on the trajectory ($\bbox{n},
\bbox{R}$),  is conveniently built from  column ``wave
functions'' $\phi $, $\phi = {u \choose v}$ which satisfy the  equation
\begin{equation}
\left(iv{\partial\over{\partial x}}+\hat{H}^{R}(x)\right) \phi=0 \; ,
\label{xra}
\end{equation}
here  
$\hat{H}^{R}(x)$ stands for
$\hat{H}_{\varepsilon,\bbox{n}}^{R}(\bbox{r})$ at the
trajectory point 
$\bbox{r}= x\bbox{n} + \bbox{R}$.

Denote $\bar{\phi }$ the row built from a column $\phi $ by the
following rule:
\[
\bar{\psi } \equiv \psi^{T} \tau_{y} {1\over i}\;\; \Rightarrow  \;\;   
\overline{
\left(
\begin{array}{c}
  u     \\
   v
\end{array}
\right)}
= 
\left(
\begin{array}{ccc}
  v & ,& -u
\end{array}
\right)\; .
\]

By virtue of the identity $  
\left(\hat{H}^{R}\right)^{T}= - \tau_{y} \hat{H}^{R} \tau_{y}$ ,
the row $\bar{\phi }(x)$ built from a solution to Eq.\ (\ref{xra}),
satisfies
the conjugated equation
\begin{equation}
\bar{\phi }(x) \left(-iv{\partial\over{\partial
x}}+\hat{H}^{R}(x)\right)= 0 \; .
\label{z4a}
\end{equation}

It follows from Eq.\ (\ref{xra}) combined with (\ref{z4a}), that
\begin{equation}
{d \over{dx}} \left(\bar{\phi_{a}} \phi_{b}\right) =0 \; .
\label{1ra}
\end{equation}
This relation is valid for any pair of solutions $\phi_{a}$ and $\phi_{b}$.

The Green's function is built of
the regular solutions to Eq.\ (\ref{xra}), {\it i.e.}  solutions
satisfying the following boundary conditions
\begin{equation}
\begin{array}{c}
\phi_{+}(x) \rightarrow 0 \;\; , \;\;  x \rightarrow + \infty \; ,   \\
\phi_{-}(x) \rightarrow 0 \;\; , \;\;  x \rightarrow - \infty \; .
\end{array}
\label{3ra}
\end{equation}
Denote $\phi_{\pm}^{(N)}$ the normalized solutions,
\begin{equation}
\overline{\phi}_{-}^{(N)}(x) \; \phi_{+}^{(N)}(x) =1 \;.
\label{8ra}
\end{equation}
The normalization is possible because the l.h.s. is a
(finite) constant  as it is seen from Eq.\ (\ref{1ra}).

The Green's function  reads
\begin{equation}
\gR(x_{1}, x_{2})=
\left\{
\begin{array}{rcr}
\phi_{+}^{(N)}(x_{1})\; \ofi_{-}^{(N)}(x_{2})&\;\;,\;\;&  x_{1}> x_{2}  \;;\\
\phi_{-}^{(N)}(x_{1})\; \ofi_{+}^{(N)}(x_{2})  &\;\;,\;\;&  x_{1}<
x_{2}     \;. 
\end{array}
\right.
\label{7ra}
\end{equation}
Indeed, it satisfies Eq.\ (\ref{vra1}) and Eq.\ (\ref{vra}) at $x_{1}\neq
x_{2}$, and is regular at $|x_{1}-x_{2}|\rightarrow \infty $ by virtue
of Eq.\ (\ref{3ra}).  Besides, the normalization in Eq.\ (\ref{8ra})
ensures that the discontinuity at $x_{1}= x_{2}$,
\begin{equation}
\gR(x+0,x)-\gR(x-0, x)= \hat{1} \; ,
\label{34a}
\end{equation}
has the value required 
by the $\delta$-function source in Eqs.(\ref{vra1}),
 and (\ref{vra}).

\subsection{1-point Green's function }\label{1point}

To find observables like the electric current or charge density, one
needs the Green's functions with coinciding spatial arguments, {\it i.e.} 
the 1-point Green's function.

The 1-point Green's   functions, 
$\gR_{\pm}(x)= \gR(x\pm 0, x)$, 
can be expressed via the normalized waves (see Eq.\ (\ref{7ra}))
\begin{equation}
\gR_{+}(x) = 
\phi_{+}^{(N)}(x) \ofi_{-}^{(N)}(x)
\;\; , \;\;  
\gR_{-}(x) = 
\phi_{-}^{(N)}(x) \ofi_{+}^{(N)}(x)
\; .
\label{44a}
\end{equation}

This expression can be identically written in the form,
\begin{equation}
\gR_{+}(x) = 
{1\over \bar{\phi_{-}}(x)\phi_{+}(x)} 
\phi_{+}(x) \ofi_{-}(x)
\;\; , \;\;  
\gR_{-}(x) = 
{1\over \bar{\phi_{-}}(x)\phi_{+}(x)} 
\phi_{-}(x) \ofi_{+}(x)
\; ,
\label{s6a}
\end{equation}
where the normalization of the wave functions $\phi_{\pm}$
is arbitrary. 

Note the projecting properties:
\begin{equation}
\gR_{\pm}\gR_{\pm}=\pm\gR_{\pm}\;\; ,\;\; 
\gR_{\pm}\gR_{\mp} = 0 \;\; , \;\;
{\rm Sp} \; \gR_{\pm} = \pm 1  \; .
\label{54a}
\end{equation}

Tagging electron- and hole-like excitations in accordance with the
direction of their propagation ($\pm x$ directions) and considering
examples, e.g. the normal state, one concludes that $\gR_{+}$ can be
identified as the (quasi)electron part of the Green's function, and
$\gR_{-}$ is the (quasi)hole one (and vice versa for
$\hat{g}^{A}_{\pm}$).

Denoting
\begin{equation}
a\equiv {u_{-}\over v_{-}}\;\; , \;\;  
b \equiv  {v_{+}\over u_{+}}, 
\label{dsa}
\end{equation}
where $u_{\pm}$ and $v_{\pm}$ are the components of 
$\phi_{\pm}$,
\[
\phi_{\pm}(x)= \Hcolumn{u_{\pm}(x)}{v_{\pm}(x)}
\;\; ,
\]
Eq.\ (\ref{s6a}) becomes
\begin{equation}
\gR_{+} =  {1\over 1-ab}\Hcolumn{1}{b}\Hrow{1}{-a}
\;\; , \;\;  
\gR_{-} =  
{1\over 1-ab}\Hcolumn{a}{1}\Hrow{b}{-1} \; .
\label{94a}
\end{equation}

As shown in Ref.\ \onlinecite{She85}, the 1-point  Green's 
function of the quasiclassical theory (``$\xi-$integrated''), 
$\gR$, is given by
\begin{equation}
\gR = \gR_{+} + \gR_{-} \; .
\label{04a}
\end{equation}
{\it i.e.}  
\begin{equation}
\gR =
\phi_{+}^{(N)}\ofi_{-}^{(N)}+ \phi_{-}^{(N)} \ofi_{+}^{(N)}
\;\; .
\label{w5a}
\end{equation}

In terms of $\gR$,
\begin{equation}
\gR_{\pm} = {1\over 2} \left(\gR \pm 1 \right)
\; ,
\label{a5a}
\end{equation}
and  the relations in Eq.\ (\ref{54a})
lead to the well-known normalization condition
\[
\left(\gR \right)^{2} = \hat{1} \; 
\]
and 
\[
{\rm Sp}\; \gR = 0 \;\; .
\]

Combining Eqs.(\ref{04a}) and (\ref{94a}), one gets
\begin{equation}
\gR= 
{1\over 1- ab}
\left(
\begin{array}{lr}
 1 + ab  &  -2 a   \\
 2b  & - (1 + ab)
\end{array}
\right) \; .
\label{fsa}
\end{equation}

This parameterisation of the Green's function has been recently
suggested by Schopohl and Maki\cite{SchMak95} (see, also Ref.\
\onlinecite{Sch98}). The present derivation leads quite naturally to
this decomposition, and clearly shows the physics behind it.  Seeing
that $a$ and $b$ may be interpreted as the ``local'' amplitudes of the
Andreev reflection for electron and hole (see below), we call them the
Andreev amplitudes.  The amplitudes $a$ and $b$ obey the nonlinear
Riccati equation\cite{SchMak95,Sch98}.  From the Bogoliubov - de
Gennes equation for $\psi={u\choose v}$, the Riccati equation for the
ratio $u/v$ has been derived by Nagato {\it et al}\cite{Nagato93}.

\subsection{Knot matching conditions}\label{knot}

In the quasiclassical picture, particles move on smooth  trajectories,
usually, straight lines characterised by the direction of velocity
$\bbox{n}$ (and the initial position $\bbox{R}$).
On interfaces, where potentials change on atomic scales, the
quasiclassical condition is violated,  and the quasiclassical wave
function spreads from the original trajectory to those coupled by
quantum scattering. Following Ref.\ \onlinecite{SheOza00}, we call  a
``knot'' the point where classical trajectories are tied together.

In a general case, the knot ties together $N$ in- and $N$
out-trajectories.  The in-trajectories (or channels) are those which
have the the Fermi velocity directed towards the knot; the
out-trajectory is characterised by the velocity directed from the
knot.  The in- and out-trajectories are somehow numbered, 
$l= 1,\ldots, N$; we mark the outgoing trajectories with $\prime$ so
that $k'$ stands for the k-th outgoing channels.

On the quasiclassical scale $\sim v_{F}/ \Delta $, the knot is
point-like, and one can define
the knot value of the trajectory wave function. Denote $\psi_{i}$
the 2-component wave function on the $i$-th in-coming trajectory, $i=
1,\ldots ,N$ at the point where it enters the knot, and analogously
$\psi_{k'}$ the knot value on the $k-$th outgoing trajectory.

The knot matching conditions suggested in Ref.\ \onlinecite{SheOza00} read
\begin{equation}
\psi_{k'} = \sum\limits_{i=1}^{N} S_{k'i}\psi_{i} 
\; \; ,
\; \; 
\ops_{k'} = \sum\limits_{i=1}^{N} S^{*}_{k'i}\ops_{i} \;\; , 
 \label{wsa}
\end{equation}
where $S_{k'i}$ are the elements of the unitary scattering matrix.  In
the spirit of the quasiclassical theory, $S_{k'i}$ is the normal metal
property taken at the Fermi surface; it is an electron-hole scalar.
This relation generalises the matching conditions of Ref.\ \onlinecite{She80b}
to the many channels case.

Eq.\ (\ref{xra}) together with the matching conditions in
Eq.\ (\ref{wsa}) allows one to find the 2-component amplitudes on 
 trajectories with knots, and, therefore, the Green's functions.

\subsection{Matching Green's functions}\label{match}

First consider  an isolated knot  mixing
semi-infinite trajectories
(with no more knots on them). 
With the origin chosen at the knot, the trajectory coordinate $x_{n}$
extends from $-\infty $ to $0$ on the $n$-th incoming trajectory, and
$0< x_{k'} < \infty $ on the $k'$-outgoing one. 
As before, the requirement,
\begin{equation}
\phi_{-, m}(-\infty ) = 0
\;\; , \;\;  
\phi_{+, k'}(\infty ) = 0 
\; \; , m,k=1,\ldots,N \; ,
\label{z7a}
\end{equation}
uniquely  defines the solutions $\phi_{-,n}(x_{n})$ and
$\phi_{+,k'}(x_{k'})$ (up to a normalization factor). 
The knot values of the regular solutions are conveniently written as  
\begin{equation}
\phi_{-,m}(x_{m}=0) = {\displaystyle a_{m}\choose 1}
\;\; , \;\;  
\phi_{+,k'}(x_{k'}=0) = {\displaystyle 1 \choose b_{k'}}
\; \; , m,k=1,\ldots,N \; ;
\label{27a}
\end{equation}
the parameters $a_{m}$ or $b_{k'}$ are ``bulk'' properties
independent on the knot.

The problem  is to find  the knot values
\begin{equation}
\phi_{+,l}(x_{l}=0) \equiv {\displaystyle 1 \choose b_{l}} 
\;\; , \;\;   
\phi_{-,n'}(x_{n'}=0) \equiv {\displaystyle a_{n'}\choose 1}, 
\;\; , \;\;  l,n=1,\ldots,N \; ,
\label{t3b}
\end{equation}
needed to evaluate
$\phi_{+,l}(x_{l}<0)$ and $\phi_{-,n'}(x_{n'}>0)$ and, therefore, the
Green's functions on the trajectories tied by the knot. On each of the
trajectories, the Green's
function  in the immediate vicinity to the knot is expressed via the
corresponding pair $a_{l}$ and $b_{l}$ or $a_{n'}$ and $b_{n'}$ by the
relation in Eq.\ (\ref{fsa}).

As has been shown in Ref.\ \onlinecite{SheOza00}, the  boundary condition 
for the Andreev amplitudes $a_{n'}$ and $b_{l}$ can be conveniently
formulated in terms of the determinant
\begin{equation}
{\cal D}(\{a\},\{b\})=
\det
\left|
\left|
1 -
\hat{S}\hat{a} \hat{S}^{\dagger}\hat{b}
\right|
\right|
\; ,
\label{87aa}
\end{equation}
built of the S-matrix  and the diagonal matrices $\hat{a}=
{\rm diag}(a_{1},a_{2},\ldots,a_{N})$ and $\hat{b}=
{\rm diag}(b_{1'},b_{2'},\ldots,b_{N'})$; $a_{m}$ and $b_{k'}$ entering
Eq.\ (\ref{27a}) are  
bulk parameters insensitive to the presence of the knot.

An important property of ${\cal D}$ is that it is a linear function of
any of $a$'s and $b$'s, so that identically 
\begin{equation}
{\cal D} = {\cal D}_{0}^{(l)} + a_{l}{\cal D}_{1}^{(l)}
\;\; , \;\;  
{\cal D} = {\cal D}_{0}^{(n')} + b_{n'}{\cal D}_{1}^{(n')}
\;\; , \;\;  
n',l = 1,2,\ldots,N  \; ,
\label{w3b}
\end{equation}
where 
\begin{equation}
{\cal D}_{0}^{(l)}= {\cal D}|_{a_{l}=0} \;\; , \;\;  
{\cal D}_{1}^{(l)}= {\partial\over{\partial a_{l}}}{\cal D}\;\; ,
{\cal D}_{0}^{(n')}= {\cal D}|_{b_{n'}=0} \;\; , \;\;  
{\cal D}_{1}^{(n')}= {\partial\over{\partial b_{n'}}}{\cal D}\;\; ,
\label{v3b}
\end{equation}
Also, ${\cal D}$ is a sum each terms of which is a product of an
equal number of $a$'s and $b$'s. From this one concludes that
\begin{equation}
\sum\limits_{l} {\cal D}_{0}^{(l)}=
\sum\limits_{n} {\cal D}_{0}^{(n')} \; .
\label{a6b}
\end{equation}
These identities are useful for transforming various expressions.

Rephrasing the procedure described in Ref.\ \onlinecite{SheOza00}, the knot
values of the Andreev amplitudes in Eq.\ (\ref{t3b}) can be found by the
following formulae:
\begin{equation}
  a_{n'} =   - {{\cal D}_{1}^{(n')}\over {\cal D}_{0}^{(n')}} 
\quad,\quad 
  b_{l} =   -  {{\cal D}_{1}^{(l)}\over {\cal D}_{0}^{(l)}} \; .
\label{u3b}   
\end{equation}
For the  case $N=2$, equivalent relations have been derived in
Ref.\ \onlinecite{Esc00}. 

Now, one is able to build the Green's functions Eq.\ (\ref{94a}) on
the $l$-th in- and $n$-th out trajectories.
\begin{equation}
\hat{g}_{+}^{(l)}=  {1\over {\cal D}} 
\Hcolumn{{\cal D}_{0}^{(l)}}{- {\cal D}_{1}^{(l)}}\Hrow{1}{-a_{l}}
\;\; , \;\;
\hat{g}_{-}^{(n')}=
{1\over {\cal D}}\Hcolumn{- {\cal D}_{1}^{(n')}}{ {\cal
D}_{0}^{(n')}}\Hrow{b_{n'}}{-1} \; .
\label{z3b}
\end{equation}

With the help of the identities in Eq.\ (\ref{w3b}) and Eq.\
(\ref{a5a}), one can present Eq.\ (\ref{z3b}) in the form,
\begin{equation}
\gR{}^{(l)} =
\left(
\begin{array}{lr}
2 \gamma_{l}-1&   - 2 \gamma_{l} a_{l}\\
 - {2\over a_{l}}(1 - \gamma_{l})  & -(2 \gamma_{l} -1)
\end{array}
\right)
\;\; , \;\;  
\gR{}^{(n')}=
\left(
\begin{array}{lr}
 2\gamma_{n'} - 1  & {2 \over b_{n'}} (1 - \gamma_{n'})   \\
 2 \gamma_{n'} b_{n'}  & -( 2\gamma_{n'} - 1)
\end{array}
\right) \; ,
\label{23b}
\end{equation}
where
\begin{equation}
\gamma_{l}= {{\cal D}_{0}^{(l)}\over {\cal D}}
\;\; , \;\;  
\gamma_{n'}= {{\cal D}_{0}^{(n')}\over {\cal D}} \; .
\label{33b}
\end{equation}

It follows from  Eq.~(\ref{a6b}) that
\begin{equation}
\sum\limits_{l}\left(\gR{}^{(l)} \right)_{11}
=
\sum\limits_{n}\left(\gR{}^{(n')} \right)_{11} \;.
\label{b6b}
\end{equation}
This identity is directly related to the current conservation.

Summarising, the Green's functions on trajectories linked by a knot are
calculated as follows.  First, one solves Eq.\ (\ref{xra})
with the boundary condition in Eq.\ (\ref{z7a}) on each of the
trajectories and calculates ``bulk'' functions $\phi_{-,m}(x<0)$ and
$\phi_{+,k'}(x>0)$. By this, one finds
 $a_{m}$ and $b_{k'}$ in Eq.\ (\ref{27a}). For a knot characterised by
a scattering matrix $S$, one is then able to find the determinant
${\cal D}$ Eq.\ (\ref{87aa}). 
The next step is to calculate the knot values of $b$'s on the
incoming trajectories and $a$'s on the outgoing ones using
Eq.\ (\ref{u3b}).  
Then, one solves Eq.\ (\ref{xra}) for $\phi_{+,l}(x<0)$ on
the incoming trajectories and $\phi_{-,n'}(x>0)$ on the outgoing
ones with the boundary condition $\phi_{+,l}(0)= {1 \choose b_{l}}$ and
$\phi_{-,n'}(0)= {a_{n'}\choose 1}$, respectively. 
The 1-point Green's function is then built from $\phi_{\pm}$ using the
representation in
Eq.\ (\ref{w5a}). The knot values of the Green's functions can be also
found from Eqs.(\ref{z3b}), or (\ref{23b}).

\begin{figure}[h] 
 \centerline{\epsfig{file=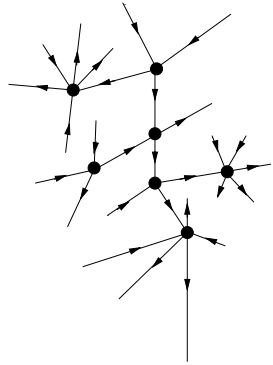,height=100pt,angle=-90}}
\caption{%
An example of a tree-like trajectory.  Pieces of the straight
lines show the trajectories before or after they enter a knot (filled
circles), 
{\it i.e.} before or after a  collision with an interface.
There is only one path connecting any two points on the tree so that
the tree is effectively 1-dimensional.
} 
\label{3-Fig} 
\end{figure}

This scheme is also applicable when the trajectories connected by the
knot under consideration may enter other knots, in a tree-like
trajectory (see Fig.\ \ref{3-Fig}). 
 As a matter of
principle, one assumes that the system under consideration is finite,
and it is surrounded by a ``clean''material where trajectories are
infinite lines without knots.  Then, one solves the problem for the
knots on the boundary and moves inwards towards the knot of interest.
In the one-dimensional topology of the tree with only one path
connecting any two knots, the procedure is unique.

\section{SANDWICH: QUASICLASSICAL GREEN'S FUNCTION} \label{sec:quasiGreen}

\begin{figure}[h] 
\centerline{
\epsfig{file=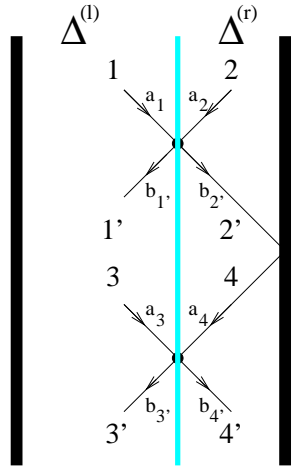,height=180pt}
}
\caption{
The trajectories of a particle in a two layer system.
}
\label{fig:sandtree}
\end{figure}

In this section we calculate the
quasiclassical Green's function in a double layer system depicted on
Fig.\ \ref{sandwich}.
The real space classical trajectory of a particle in a two layer
system is formed by multiple reflections on the outer surface and the
interface between the layers (see Fig.\ \ref{fig:sandtree}.). 

The Green's function  can found using Eq.\ (\ref{23b}). In this simple
case  with only two in- and out-trajectories,
determinant ${\cal D}$ reads
\begin{equation}
{\cal D}= R (1 - a_{1}b_{1'})(1- a_{2}b_{2'})
+
T(1 - a_{1}b_{2'})(1- a_{2}b_{1'})
\;,
\label{93b}
\end{equation}
where $R$ and $T$ are the reflection and transmission coefficients.

The problem is to find the
amplitudes $a_{m}$ on the incoming legs of a certain knot and
$b_{k'}$ on the outgoing legs. 
Since none of the legs lead to infinity the situation is slightly
different than in the previous chapter and the procedure has to be
modified. 

The argument used to find the equations for the amplitudes $a_{m}$
and $b_{k'}$ is the invariance with respect to translation parallel to
the interface. 
Suppose we know a transfer matrix which relates the wave function
$\phi_{-}$ in the legs say "2" and "4". Because of the translational
invariance these two functions can differ only by a 
prefactor. In another words $\phi$ is eigenfunction of the transfer
matrix and the ratio of the components of $\phi$ must be the
same in  both corresponding legs.
Thus the amplitudes $a$ in legs "2" and "4" are equal $a_{2} = a_{4}$.
One can argue in the same way
to show that $a_{1} = a_{3}, b_{1'} = b_{3'}$ and so on.

We can proceed in the following manner.
First we find the transfer matrix as a
product of knot transfer matrix and propagator across the layer.
Then we relate the amplitudes in corresponding legs. We get a set
of $4$ equations for $4$ unknowns $a_{1}, a_{2}, b_{1'}$ and $b_{2'}$
(see Fig.\ \ref{fig:sandtree}. for notations). 

The knot transfer matrix connecting the wave function in the incoming and
outgoing legs on a knot can be written as follows
\begin{equation}
{\cal M}_{2' \leftarrow 2} \propto 
\left(
\begin{array}{cc}
1 & a_{1} \\
b_{1'} & 1
\end{array}
\right) 
\left( 
\begin{array}{cc}
R & 0 \\
0 & 1
\end{array} \right)
\left(
\begin{array}{cc}
1 & a_{1} \\
b_{1'} & 1
\end{array}
\right)^{-1} \, . 
\label{Masb}
\end{equation}
The transfer matrix\cite{SheOza00} for the channels on the right side
is a function of the amplitudes only on the left side. 

The propagator across the layer is found by integration of 
Eq.\ (\ref{xra}). For homogeneous superconductor the expression reads:
\begin{equation}
{\cal U}(x) \propto \left(
\begin{array}{cc}
1 & a_{0}^{(r)} \\
b_{0}^{(r)} & 1
\end{array}
\right) 
\left( 
\begin{array}{cc}
e^{i {\xi^{(r)}x \over v_{F}}} & 0 \\
0 & e^{-i {\xi^{(r)}x \over v_{F}}}
\end{array} \right)
\left(
\begin{array}{cc}
1 & a_{0}^{(r)} \\
b_{0}^{(r)} & 1
\end{array}
\right)^{-1} \; .
\label{Mbsb}
\end{equation}
Here, 
$a_{0}^{(r)} = \Delta^{(r)} / (\varepsilon + \xi^{(r)})$ 
and
$b_{0}^{(r)}= \Delta^{(r)*} / (\varepsilon + \xi^{(r)})$ are the bulk
values of  $a$ and $b$ in the right layer,
and $\xi^{(r)} = \sqrt{\varepsilon ^2 - |\Delta^{(r)}|^2}$
(${\rm Im}\;\xi^{(r)}>0$).

Once the transfer matrices for both sides are known, one can write down
the $4$ equations for the four unknown amplitudes $a_{1,2}$ and
$b_{1',2'}$. Solving the equations and using the expressions relating
$a$'s and $b$'s in the in/out going channels from the previous chapter
one finds the amplitudes in every leg. Thus the Green's function can
be constructed and the density of states calculated.

When the numerical calculations are carried out, it is easier to use an
iterative procedure instead of solving the set of four nonlinear
equations.  One starts with the bulk values of $a$ and $b$ in, say,
right side and calculates the corresponding amplitudes on the
left. These values are then used to calculate back $a$ and $b$ on the
right side. This process is continued until self-consistency is
reached.

\section{SANDWICH: EXACT GREEN'S FUNCTION}  \label{sec:exactGreen}

In this section we calculate the exact Green's function for a sandwich.

Consider a sandwich formed by two
homogeneous layers with thicknesses $a^{(l,r)}$ and order
parameters $\Delta^{(l,r)}$; here and below, the superscript $(l)$ and
$(r)$ label quantities which refer to the left and right layer,
respectively, see Fig.\ \ref{sandwich}(a).

The left (right) layer occupies the region  $- a^{(l)}<x<0$ ($0<x<a^{(r)}$).
The interface is  characterised by the scattering $S$-matrix,
\begin{equation}
S = \left(
\begin{array}{lr}
t   &  r  \\
 r  & t
\end{array}
\right) \; ,
\label{83b}
\end{equation}
with the transmission and
reflection amplitudes $t$ and $r$ for the case of $\delta$-function
interface potential.
 The outer walls are impenetrable.
The goal is to find the exact Green's function
$G(\bbox{r},\bbox{r}')$ from the Gor'kov
equation.

Due to the translational invariance in the direction parallel to
the layers, the problem is effectively one-dimensional. 
For a given value of the parallel momentum $\bbox{p}_{||}$, the
Green's function, $\hat{G} =
\hat{G}_{\varepsilon, \bbox{p}_{||}} (x,x')$,
 depends on the 
two coordinates $x$ and $x'$
as well as the energy
$\varepsilon$.
The Gor'kov equation reads:
\begin{equation}
(\varepsilon - \hat{H}) \hat{G}(x, x') =  \hat{1} \delta (x- x')
\quad ; \quad
\hat{H} = 
\left(
\begin{array}{cc}
\hat{\xi} & \Delta \\
\Delta^{*} & - \hat{\xi}
\end{array}
\right) \; .
\label{Moob}
\end{equation}
For  the parabolic spectrum,  $\hat{\xi} =
{\hat{\bbox{p}}^{2}/ 2m} - {p_{Fx}^{2}/ 2m}$, where $p_{Fx}$ is 
$x$-projection of the Fermi momentum $p_{Fx}^{2} = p_{F}^{2} -
{\bbox{p}}_{||}^{2}$. The Green's function is
continuous at $x=x'$: $\hat{G}(x^{+},x) = \hat{G}(x^{-},x)$.
Its derivatives suffer a jump generated by the delta-function on the
r.h.s. of Eq.\ (\ref{Moob}): 
\begin{equation}
\hat{p}_{x} \left.\hat{G}(x,x')\right|^{x=x'+0}_{x=x'-0} = 2m/i \hbar
\; .
\label{Moob1}
\end{equation} 
The boundary conditions at $x=0$ corresponding to the semi-transparent
interface 
are conveniently written as
\begin{equation}
\hat{P}^{\sigma} \hat{G}(0^{+},x') = \sum \limits_{\sigma' = \pm}
{\cal M}_{\sigma \sigma'} \hat{P}^{\sigma'} \hat{G}(0^{-},x') \quad
\; \sigma = \pm \; ,
\label{M2qb}
\end{equation}
vhere the projectors $\hat{P}^{\pm}$,
\begin{equation}
\hat{P}^{\pm} = {1 \over 2} (1 \pm \hat{p}_{x} /p_{Fx})
  \;\;,
\label{43b}
\end{equation}
and 
the transfer matrix ${\cal M}$,
\begin{equation}
{\cal M} = \left(
\begin{array}{cc}
1/t^{*} & r/t \\
r^{*}/t^{*} & 1/t
\end{array}
\right) \; .
\label{M3qb}
\end{equation}

It is convenient to
express the Green's function in terms of two-component functions
$\Phi_{\nu,\sigma}(x)$ and $\overline{\Phi}_{\nu,\sigma}(x')$ defined
as follows:
\begin{equation}
\Phi_{\nu \sigma}(x) = \phi_{\nu} e^{i \sigma p_{\nu} x}
\quad ; \quad 
\overline{\Phi}_{\nu \sigma}(x') = \overline{\phi} e^{-i \sigma
p_{\nu} x'} \; .
\label{M4qb2}
\end{equation}
Here $\sigma=\pm$, $p_{\nu} = p_{Fx} + \nu \xi /m$ is the electron
 ($\nu=+$) or hole ($\nu=-1$)  momentum,
and  $\phi_{\nu}$ is the solutions to
matrix equation 
\begin{equation}
\left(
\begin{array}{cc}
\nu \xi & \Delta \\
\Delta^{*} & -\nu \xi 
\end{array} 
\right) 
\phi_{\nu} = \varepsilon \phi_{\nu} \; ,
\end{equation}
and $\overline{\phi}_{\nu}$ is the adjoint row-vector defined in such
a way that $\overline{\phi}_{\nu} \phi_{\mu} = \delta_{\nu \mu}$.

The Green's function can be then written
as:
\begin{equation}
i \hbar v_{F_{x}} \hat{G}(x,x') \sigma_{z}= \left\{
\begin{array}{lc}
\sum \limits_{\nu,\nu',\sigma,\sigma'} \Phi_{\nu,\sigma}(x) \langle
\nu,\sigma | Q^{>} | \nu' \sigma' \rangle \overline{\Phi}_{\nu',\sigma'}
(x') & \mbox{ for } \; x > x' \\
\sum \limits_{\nu,\nu',\sigma,\sigma'} \Phi_{\nu,\sigma}(x) \langle
\nu,\sigma | Q^{<} | \nu' \sigma' \rangle \overline{\Phi}_{\nu',\sigma'}
(x') & \mbox{ for } \; x < x'
\end{array}
\right. \; .
\end{equation}
The $4 \times 4$ matrices  $\hat{Q}^{<,>}$  are constants.
It has different form for $x > x'$ and $x < x'$. 

Inserting the Green's function into Eqs.(\ref{Moob}-\ref{Moob1}) and
using the matching condition at the interface in Eq.\ (\ref{M2qb}) one
gets $\langle \nu, \sigma | Q^{<,>}|
\nu' \sigma' \rangle$. 
After straightforward but lengthy calculations, we have found
the following rather simple result:
\begin{eqnarray}
\!\!\!\!\!
\left(
\begin{array}{cc}
\!\!\!
\tau_{z} Q^{>}_{11} &\!\!\! \tau_{z} Q^{>}_{12}  \\
\!\!\!
\tau_{z} Q^{>}_{21} &\!\!\! \tau_{z} Q^{>}_{22} 
\end{array}
\right) 
\!\!\!
&=&  
\!\!\!\!
{1 \over D} \left(
\begin{array}{cc}
\!\!\!
w_{+} - A_{+} A_{-} w_{+} w_{-} &
\!\!\!
A_{-} w_{+} (w_{-} - 1) \\
\!\!\!
A_{+} w_{-} (w_{+}-1) &
\!\!\!
w_{-} - A_{+} A_{-} w_{-} w_{+}
\end{array} \right) \; ,
\label{M4qb} \\ 
D &=& 1 - A_{-} A_{+} \mbox{Tr} (w_{+} w_{-}) \; .
\end{eqnarray}
Here,  $w_{\pm}$ is a
$2\times2$ matrix, $w_{\pm} = W_{\pm} /\mbox{Tr} W_{\pm}$,
\begin{equation}
W_{\nu} = \left( 
\begin{array}{cc}
1 \\
e^{2i p_{\nu}^{(r)} a^{(r)}}
\end{array}\right) \cdot
\left( 
r^{*} + |r|^{2} e^{2i p_{\nu}^{(l)} a^{(l)}}  , 
|r|^{2} + r  e^{2i p_{\nu}^{(l)} a^{(l)}}
\right) \;\;, 
\label{53b}
\end{equation}
and $A_{\pm}$ are 
(scalar) coefficients,
\begin{equation}
A_{\pm} = {\overline{\phi}_{\mp}^{(r)} \phi_{\pm}^{(l)} \over
\overline{\phi}_{\pm}^{(r)} \phi_{\pm}^{(l)}} \; ,
\label{63b}
\end{equation}
which have the physical meaning of the  Andreev reflection amplitudes
for a  transparent interface. 

The Gor'kov Green's function for a sandwich is known from the literature
\cite{AshAoyHar89}.  Eq.\ (\ref{M4qb}) agrees with the previous work,
giving simple and concise form for the Green's function.

\subsection{Coarse-grained Green's function}

As have been already mentioned in the Introduction, the quasiclassical
theory does not attempt to give any good description on the truly
microscopically local level. Instead, it supplies knowledge about
coarse-grained observables.  Accordingly, one should derive the
coarse-grained Green's function corresponding to the exact theory before
comparing it with the quasiclassical counterpart.

\begin{figure}[h]
\centerline{\epsfig{file=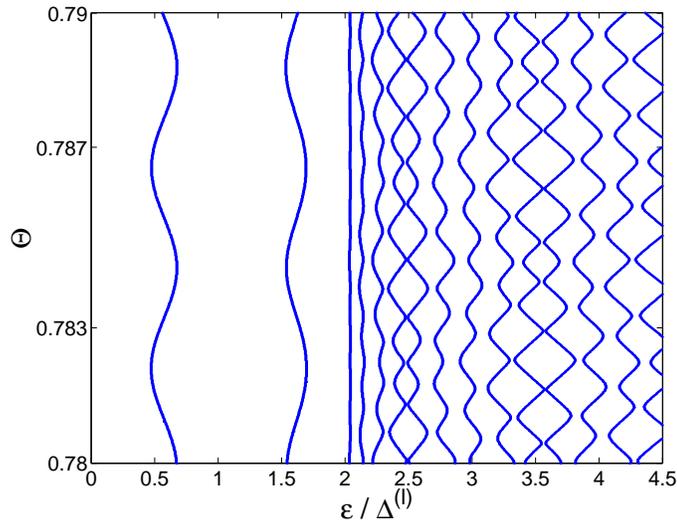,height=200pt}}
\caption{
Discrete energy levels in then sandwich Fig.\ \ref{sandwich}(a)
corresponding to eigenstates with a definite momentum $ p_{||}$.
The curves show how the position of the level $\varepsilon $ changes
under small variations of the angle $\Theta $, $\cos \Theta =
p_{||}/p_{F}$. The sandwich with 
$a ^{(l)}= v_{F}/\Delta^{(l)}$ is thick on the scale of the Fermi wave
length: $a^{(l)}=1000/p_{F}$; the transparency of the interface
$T=0.9$.  Other parameters of the sandwich: $a ^{(r)}= 3 a^{(l)}$, 
$\Delta ^{(r)}= - 2 \Delta ^{(l)}$.}
\label{fig:BS1}
\end{figure}

The space averaging of the partial ({\it i.e.} momentum resolved)
density of states is rather simple. For a given $\bbox{p}_{||}$, the
density of states is proportional to  the imaginary part
$G_{\bbox{p}_{||}}(x,x' \rightarrow x)$. It is seen from
Eq.\ (\ref{M4qb2}) that the elements of the $Q$-matrix which are
off-diagonal in $\sigma $, create  rapidly oscillating
terms. Therefore, space averaging amounts to ignoring the 
off-diagonal terms.

Even reduced to a low spatial resolution, the Green's function remains
to be a fast function of the momentum direction. For instance, the
density of states is non-zero only at those {\em discrete} values of
$\bbox{p}_{||}$ where the energy variable $\epsilon $ equals to the
energy of the bound state for the finite motion in the $x$-direction.
Fig.\ \ref{fig:BS1} shows the fast periodic motion of the energy level
positions under a small variation of $p_{||}$.  An integration within
a small interval of the angles is required to transform the density of
states from a sequence of the $\delta $-function spikes into a smooth
function of the momentum direction.

\begin{figure}[h]
\centerline{\epsfig{file=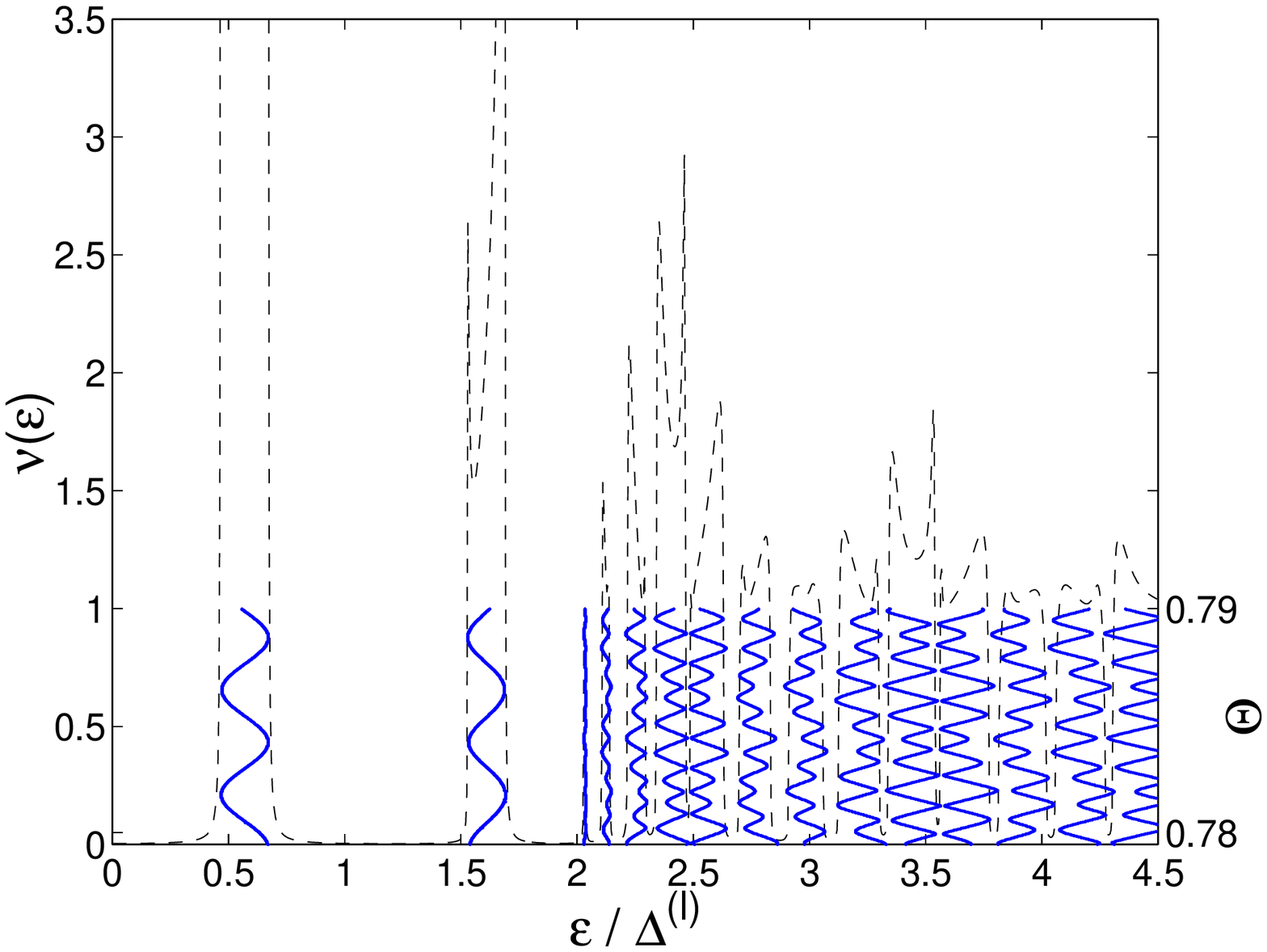,height=200pt}}
\caption{
The energies of the bound states as a function of angle (full line)
superimposed on the quasiclassical angular resolved density of states
(dashed line). The parameters of the double layer are
$\Delta^{(r)}=-2\Delta^{(l)} , a^{(l)}= v_{F}/\Delta^{(l)} , a^{(r)}=
3 v_{F} / \Delta^{(l)} , T=0.9, \Theta={\pi/4}$.  } \label{fig:QCxBS1}
\end{figure}

In the next section we compare the coarse-grained density of states
extracted from the exact Green's function with the quasiclassical
trajectory resolved DOS.

\section{DOS: QUASICLASSICAL VS. EXACT} \label{sec:sandwich}

In the most simple case where the order parameter is the same in the
left and right layers, the both theories reproduce the BCS density of
states for any transparency of the interface (as a consequence of the
Anderson theorem).  Nontrivial comparison requires an inhomogeneous
order parameter.

First, we analyze a generic ``sample'', parameters of which are in no
special relation to each other. In Fig.\ \ref{fig:QCxBS1}, the energy
of the bound states of the exact theory are superimposed on the
quasiclassical DOS.

\begin{figure}[h]
\centerline{\epsfig{file=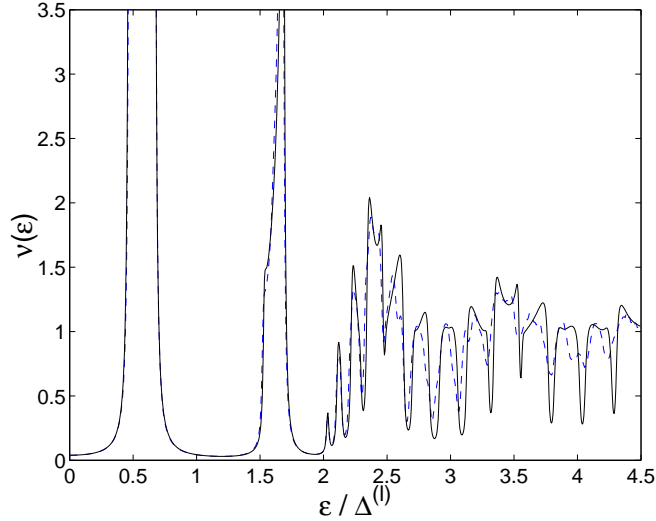,height=200pt}}
\caption{
The angular resolved density of states calculated using both exact 
(dashed line) and quasiclassical (full line)
methods. The parameters of the double layer are
$\Delta^{(r)}=-2\Delta^{(l)} , a^{(l)}= v_{F}/\Delta^{(l)} , a^{(r)}=
3 v_{F} / \Delta^{(l)} , T=0.9, \Theta={\pi/4}$.
}
\label{fig:QCxBdG1}
\end{figure}

One sees that indeed the periodic motion of the levels occurs in the
allowed bands predicted by the quasiclassics.  From Fig.\
\ref{fig:QCxBdG1}, which shows the DOS as a function of energy, one
may conclude that the quasiclassical theory gives correct energy
dependence, excepting perhaps some very fine details.

\begin{figure}[h]
\centerline{\epsfig{file=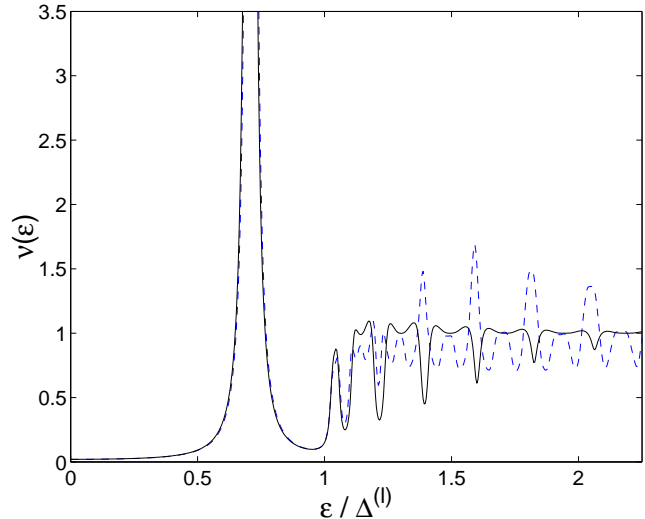,height=200pt}}
\caption{
The quasiclassical (full line) and exact (dashed line) angular
resolved density of states for $\sqrt{NOT}$ case. The 
parameters of the double layer 
are 
$\Delta^{(r)} = - \Delta^{(l)}, a^{(l)}=4 v_{F} / \Delta^{(l)} ,
a^{(r)}=4 v_{F} / \Delta^{(l)}
, T=2/3, \Theta={\pi/4}$.
}
\label{fig:sqrtNOT}
\end{figure}

In Fig.\ \ref{fig:sqrtNOT}, we show the DOS for a specially selected
``sample'' for which non-classical effects are expected to be most
pronounced.  The sample is left-right symmetric except that
$\Delta^{(l)} = - \Delta^{(r)}$. Because of  the  symmetry, 
the contribution of the loop in Fig.\ \ref{andreevABC}(c) survives  the
coarse-grained 
angular averaging. Besides, the knot scattering
matrix $S$ of the ``sample'' is intentionally chosen as
``$\sqrt{NOT}$'' {\it i.e.} 
\begin{equation}
S =  {1\over \sqrt{2}}\left(
\begin{array}{lr}
1   &  i\\
 i  & 1
\end{array}
\right)  \quad,\quad 
S^2 = i \left(
\begin{array}{lr}
0   &  1  \\
 1  & 0
\end{array}
\right) \; ,
\label{73b}
\end{equation}
In this case due to the quantum interference of reflected and
transmitted, the interface is fully transparent if the particle hits
it twice (provided the sandwich is symmetric).  The quasiclassical
theory misses this phase sensitive effect since only the probabilities
enter the theory.

\begin{figure}[h]
\centerline{\epsfig{file=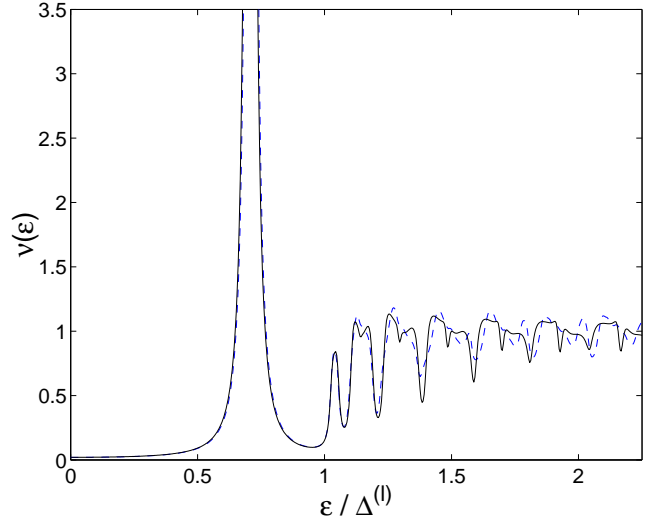,height=200pt}}
\caption{
The angular resolved density of states for ``almost'' $\sqrt{NOT}$
case. The parameters of the double layer 
are 
$\Delta^{(r)} = - \Delta^{(l)}, a^{(l)}=4 v_{F} / \Delta^{(l)} ,
a^{(r)}=4.1 v_{F} / \Delta^{(l)}
, T=2/3, \Theta={\pi/4}$.
The full line corresponds to the quasiclassical theory, the dashed
line to the exact calculation. 
}
\label{fig:sqrtNOTa}
\end{figure}

As clearly seen from Fig.\ \ref{fig:sqrtNOT}, the two approaches show
distinct result for this ``sample''.  We attribute the disagreement to the loop
contribution. This point of view is supported by
Fig.\ \ref{fig:sqrtNOTa} where DOS is shown for the sample differing
from the previous one only in a slight violation of the left-right
symmetry. For this sample, the disagreement is much less
pronounced. This can be understood since in an asymmetric sample, the
simple loop contribution averages to zero.

\section{CONCLUSIONS} \label{conclusions}

The main goal of the study has been to check the validity of the
quasiclassical scheme suggested in Ref.\ \onlinecite{SheOza00} for the
description of multiple reflections in layered superconducting
systems.  For this, we have compared the angular (trajectory) resolved
density of states obtained from the full Gor'kov equation and the
quasiclassical theory.  From the comparison, we conclude that the
approach based on the notion of quasiclassical tree-like trajectories
\cite{SheOza00} is in qualitative agreement with the exact
theory. Some quantitative disagreement which comes as no surprise,
seems to be under control. The point here is that the assumption about
the tree-like character of classical trajectories can be justified
only for rough interfaces when the system becomes non-integrable,
whatever small the disorder. However, the exact Green's function is
calculated in the ideal geometry where the trajectories have loops (as
in Fig.\ \ref{sandwich}).  We ascribe the disagreement to the loop
contribution.  Choosing the geometry, we able to control to some
extent the loop contribution. Indeed, the disagreement is most
pronounced in the symmetric ``sample'' (see Fig.\ \ref{fig:sqrtNOT})
where the contribution of the simplest loops shown in Fig.\
\ref{sandwich}(b) survives coarse-grained averaging.  A small
asymmetry, which destroys their contribution, considerably reduces the
disagreement (see Fig.\ \ref{fig:sqrtNOTa}).  Note that loops always
exist in the ideal specular geometry: even for an asymmetric sandwich,
$a^{(l)}\neq a^{(r)}$, a loop is formed after every second collision
with the interface. These higher order loops are not destroyed by
either the angular averaging or the averaging with respect to the
layer thickness as in Ref.\ \onlinecite{Nagato93} and
\onlinecite{Nag98}. In our opinion, the difficulties with the
quasiclassical theory reported in Ref.\ \onlinecite{Nagato93} and
\onlinecite{Nag98} originate in the contribution of the high order
loops.

Finally, our conclusion about the validity of the quasiclassical
theory is as follows.  The theory {\em is} applicable for the
description of a typical real sample where some roughness of the
interfaces and surfaces is inevitably present. Since real typical
samples are most probably non-integrable, the trajectories are
tree-like, and the scheme suggested in Ref.\ \onlinecite{SheOza00} is
valid. However, the scheme does {\it not} include certain physics
which may be of importance in some situations where the sample is
manufactured to enhance the role of interfering paths. Then, one
should use the full version of the Gor'kov equations for the
description of such a special system.

\end{document}